\documentclass[%
 reprint,
 amsmath,amssymb,
 aps,
]{revtex4-2}
\usepackage{xcolor}%
\usepackage{graphicx}
\usepackage{dcolumn}
\usepackage{bm}

\newcommand*\diff{\mathop{}\!\mathrm{d}}
\begin{document}

\preprint{APS/123-QED}

\title{Topological Defects as Nucleation Points of the Nematic--Isotropic Phase Transition in Liquid Crystal Shells}

\author{Yucen Han}
\email{yucen.han@strath.ac.uk}
 \affiliation{Department of Mathematics and Statistics, University of Strathclyde, Glasgow, United Kingdom}
\author{Jan Lagerwall}%
\affiliation{Experimental Soft Matter Physics group, Department of Physics and Materials Science, University of Luxembourg, L-1511, Luxembourg
}
\author{Apala Majumdar}%
\affiliation{Department of Mathematics and Statistics, University of Strathclyde, Glasgow, United Kingdom
}

\begin{abstract}
The transition from a nematic to an isotropic state in a self-closing spherical liquid crystal shell with tangential alignment is a stimulating phenomenon to investigate, as the topology dictates that the shell exhibits local isotropic points at all temperatures in the nematic phase range, in the form of topological defects. The defects may thus be expected to act as nucleation points for the phase transition upon heating beyond the bulk nematic stability range. Here we study this peculiar transition, theoretically and experimentally, for shells with two different configurations of four +1/2 defects, finding that the defects act as the primary nucleation points if they are co-localized  in each other's vicinity. If the defects are instead spread out across the shell, they again act as nucleation points, albeit not necessarily the primary ones. Beyond adding to our understanding of how the orientational order--disorder transition can take place in the shell geometry, our results have practical relevance for, e.g., the use of curved liquid crystals in sensing applications or for liquid crystal elastomer actuators in shell shape, undergoing a shape change as a result of the nematic--isotropic transition.       
\end{abstract}

\maketitle


\section{\label{sec:introduction}Introduction}
The transition between a regular isotropic liquid to a nematic liquid crystal phase with long-range orientational order is a fascinating case of spontaneous symmetry breaking which has been studied extensively, especially for bulk liquid crystals (LCs). There are different kinds of LCs - Nematic Liquid Crystals (NLCs) with long-range orientational ordering and locally preferred or distinguished averaged directions of molecular alignment (commonly known as nematic directors), Smectics (Sm) with layered structures of which smectic-A (SmA) and smectic-C are widely studied smectic phases and Cholesteric Liquid Crystals with twisted helical structures \cite{de1993physics}. For the case of strong confinement, the influence of surfaces and, when present, tortuosity can have significant impact on phase transitions in general \cite{IANNACCHIONE1992,Jin2005,Enz2009}. When topological defects are induced by the confinement, they may even change the phase sequence, e.g., inducing a localized nematic state in a SmA phase subject to normal boundary conditions in a cylinder \cite{Kralj1996}. A situation that is even more intriguing is that of liquid crystal shells, in which the LC phase is confined in a thin spherical layer between an internal droplet of immiscible isotropic liquid and a continuous phase of immiscible isotropic liquid
 \cite{lopez2011drops,urbanski2017liquid}. While the inner and outer liquids can have a variety of compositions, they are most commonly water solutions of a molecule that help to stabilize the interface and control the alignment of the standard uniaxial LC director. A fully nematic shell subject to tangential boundary conditions on both interfaces must, according to the Poincaré--Hopf theorem, exhibit a total topological defect strength of +2 on each interface, internal and external. While transitions between different LC phases have been carefully studied in shells, in particular transitions between nematic and SmA \cite{liang2011nematic,lopez2011nematic,sevc2012defect,liang2013tuning}, the nematic--isotropic (NI) transition has not been studied in detail so far, although it is actually a particularly intriguing transition in this geometry.

The confinement-induced presence of topological defects constitutes an interesting peculiarity when considering the nucleation of the transition from nematic to isotropic. While in a defect-free nematic phase, there are no obvious nucleation points, the fact that the orientational order breaks down locally within topological defects means that each defect could be considered a nucleus of the isotropic state from which the transition might be expected to grow. Particularly interesting is the common situation in tangential-aligned shells of four +1/2 disclinations, because in contrast to integer defects, they persist throughout the shell and form a defect line that connects the inner and outer interface. Most often, shells are quite asymmetric in shape due to density mismatch driving the inner isotropic droplet up or down within the LC, yielding antipodal points of minimum and maximum thickness within the shell. This geometry favors the collection of all four +1/2 disclinations near the thinnest point, to minimize the length of the defects, although it comes at the cost of increased elastic director field distortion near the defects \cite{fernandez2007novel}. By making the shell very thin on average, however, the asymmetry is reduced and now the most beneficial configuration is to spread the defects apart as much as possible to minimize the elastic distortion, in the ideal case yielding a tetrahedral distribution of the four +1/2 defects across the shell \cite{lopez2011frustrated}.

In this paper we analyze theoretically and test experimentally how the nematic-isotropic transition proceeds in a highly asymmetric and an almost symmetric shell respectively, both exhibiting four +1/2 defects in the starting configuration, all collected near the thinnest point in the former case while spread out far from each other in the latter case. We find that the defects indeed act as nucleation points, the transition clearly starting from the four defects in the asymmetric shell, with other regions nucleating isotropic domains only later in the process. In the more symmetric shell, in contrast, the transition first nucleates away from any defects, and only slightly later we see that the isotropic phase grows out from the defects as well, which thus still serve as nuclei, but not the first ones. We argue that the difference can be attributed to the greater elastic deformation cost of having all four defects collected near the thinnest point compared to distributing the defects, effectively reducing the transition temperature near the defect collection by an amount that is large enough to detect experimentally.

The paper is organised as follows. In Section~\ref{sec:experimental_details}, we give the essential experimental details of the shell manufacturing process. In Section~\ref{sec:LDG}, we review the Landau-de Gennes (LdG) continuum theory for nematic liquid crystals (NLCs) and present numerical computations of NLC equilibria on symmetric and asymmetric shells in the LdG framework, accompanied by comparisons with experimental data. These numerical computations demonstrate the accumulation of defects near the thinnest part of asymmetric shells, for stable LdG equilibria. In Section~\ref{sec:OF}, we model the NLC profile on a spherical surface, with tangential conditions, in the simplest Oseen-Frank (OF) framework. The OF framework is simpler and less detailed than the LdG framework, restricted to uniaxial NLCs with a single distinguished director and constant degrees of orientational ordering. The LdG framework can account for uniaxial and biaxial NLC phases, with multiple directors and variable degrees of orientational ordering, along with defects of all dimensionalities. We show that the simple OF model on a spherical surface can capture the essential structural details of the more sophisticated LdG equilibria on shells. In fact, we can compute semi-explicit director profiles in the OF framework and these profiles contain quantitative information about defect locations and interactions for asymmetric and symmetric shells respectively. In Section~\ref{sec:clearing}, we present experimental results on both heating and cooling transitions in tangentially anchored NLC-filled shells, both symmetric and asymmetric. The experimental data supports that the clearing temperature (for which the shell is largely isotropic) is reduced in asymmetric shells compared to their symmetric counterparts and we adapt mathematical models in \cite{mottram1997disclination, mottram2000defect} to the shell problem in the OF framework and the model reproduces the reduced clearing temperatures for asymmetric shells and the faster growth of the isotropic phase in asymmetric shells with increasing temperature, compared to their symmetric counterparts. We conclude with some remarks about the strengths and limitations of our work, along with avenues for further model development in Section~\ref{sec:conclusions}.

\section{Experimental Details}
\label{sec:experimental_details}
The shells were made from 4'-Octyl-biphenyl-4-carbonitrile (8CB) purchased from Synthon Chemicals (Germany) using a nested capillary microfluidic device constructed in-house (for details, see \cite{urbanski2017liquid}) and a Fluigent (France) MFCS pneumatic flow control device. The inner and outer isotropic phases were both isotropic aqueous solutions of polyvinylalcohol (PVA), molar mass $13-23$~kg/mol, 87--89\% hydrolyzed, at 1~wt.\% concentration. The shells were produced with 8CB heated slightly into the isotropic phase but the collection bath was at room temperature, leading to a rapid cooling of the shells into a disordered SmA state. This was the starting configuration for the further analysis with a polarizing microscope (Olympus BX-51) equipped with a Linkam T95-PE hot stage for temperature control and a Sony FDR-AXP33 camcorder for video recording. The shell suspension (with the 1~wt\% PVA solution inside and outside) was collected into a flat capillary for microscopic investigation.

\section{Numerical Computations of Landau-de Gennes Equilibria 
on shells and Comparison to Experimental Results}
\label{sec:LDG}

We perform the numerical simulations of nematic shells in the Landau-de Gennes (LdG) framework with the LdG order parameter $\mathbf{Q}$ \cite{de1993physics}. The LdG order parameter, $\mathbf{Q}$, is a macroscopic order parameter given in terms of a symmetric traceless $3\times 3$ matrix, whose eigenvectors model the nematic directors (locally preferred directions of molecular alignment) and the corresponding eigenvalues measure the degree of orientational order about the eigenvectors. 
A $\mathbf{Q}$-tensor is biaxial if $\mathbf{Q}$ has three distinct eigenvalues, uniaxial if $\mathbf{Q}$ has a pair of degenerate non-zero eigenvalues, and isotropic if $\mathbf{Q}=\mathbf{0}$ \cite{de1993physics,mottram2014introduction}. A biaxial NLC phase has both primary and secondary nematic directors, and a uniaxial phase only has a primary nematic director, such that all directions perpendicular to the uniaxial director are physically equivalent.
The simplest form of the LdG free energy functional is given by \cite{majumdar2010EJAM}
\begin{equation}\label{eq:energy_LdG}
    F(\mathbf{Q})=\int_{\Omega} \frac{L_1}{2}|\nabla\mathbf{Q}|^2+f_b(\mathbf{Q})\diff V,
\end{equation}
where the bulk energy density is 
\begin{equation}\label{eq:bulk_LdG}
    f_b(\mathbf{Q}):=\frac{A}{2}\textrm{tr}\mathbf{Q}^2-\frac{B}{3}\textrm{tr}\mathbf{Q}^3+\frac{C}{4}(\textrm{tr}\mathbf{Q}^2)^2.
\end{equation}
Here $L_1$ is an elastic constant, $A$ is a temperature-dependent constant, $B$ and $C$ are material-dependent constants, and $\Omega$ is the volume of the body under consideration.
 
 The parameter $A$ takes the role of a rescaled temperature, with three characteristic values: (i) $A = 0$, below which the isotropic phase $\mathbf{Q} = 0$ is unstable, (ii) the nematic-isotropic transition temperature, $A = B^2/27C$ (transition temperature $T_{NI}$), at which $f_B$ is minimized by the isotropic phase and a continuum of uniaxial states with the same energy, and (iii) the nematic superheating temperature, $A = B^2/24C$ (maximum clearing temperature $T^*$) above which the isotropic state is the unique critical point of $f_b$. For a given temperature $A<B^2/24C$, the bulk potential, $f_b$, has a minimiser belonging to the set of ordered uniaxial nematic states: $\mathcal{N}:=\left\{\mathbf{Q}\in \mathcal{M}^{3\times 3}: Q_{ij} = Q_{ji}, Q_{ii} = 0, \mathbf{Q} = s_+(\mathbf{n}\otimes\mathbf{n} - \mathbf{I}/3)\right\}$, where 
\begin{equation}\label{eq:s+}
s_+ = \frac{B+\sqrt{B^2 - 24AC}}{4C}
\end{equation} 
and $\mathbf{n}\in \mathcal{S}^2$ arbitrary, referred to as the uniaxial director. The elastic energy density, $|\nabla \mathbf{Q}|^2$, penalises spatial inhomogeneities and in particular, defines the elastic distortion costs associated with defects and their locations. The physically observable configurations are modelled by local or global energy minimisers, subject to the imposed boundary conditions.

The shell domain is denoted by, $\Omega = B(\mathbf{0},R_o)\backslash B((0,0,\delta),R_i)$, where $B((0,0,\delta),R_i)\subset B(\mathbf{0},R_o)$ and $\delta$ is the eccentricity: the distance between the inner
and outer spherical centers. $R_o$ and $R_i$ are the radii of the outer and inner spherical interfaces respectively.

We nondimensionalize the system using the following rescaling 
\begin{gather}
\bar{\mathbf{x}} = \mathbf{x}/R_o, \ \bar{\mathbf{Q}} = \sqrt{\frac{27C^2}{2B^2}}\mathbf{Q},\  \bar{\mathcal{F}} = \frac{27C^3}{2B^4R_o^3}\mathcal{F}.
\end{gather}
Dropping all bars for convenience, the dimensionless LdG functional can be written as
\begin{equation}\label{eq:energy-3d}
    \mathcal{F}(\mathbf{Q})= \int_{\Omega} \left\{\frac{\xi_R^2}{2}|\nabla\mathbf{Q}|^2+\frac{t}{2}\textrm{tr}\mathbf{Q}^2-\sqrt{6}\textrm{tr}\mathbf{Q}^3+\frac{1}{2}(\textrm{tr}\mathbf{Q}^2)^2\right\}\diff\mathbf{x},
\end{equation}
with non-dimensionalised domain $\Omega=B(\mathbf{0},1)\backslash B(\mathbf{c},\rho)$, where $\rho = R_i/R_o$, $\mathbf{c} = (0,0,\delta)/R_o=(0,0,c)$, satisfying $c > \rho > 0$ and $c+\rho < 1$, the reduced temperature $t = \frac{27AC}{B^2}$, and $\xi_R = \sqrt{\frac{27CL_1}{B^2R_o^2}}$.

The planar degenerate or tangential anchoring can be imposed by adding the surface energy
\begin{equation}\label{eq:tangential}
F_s = \int_{\partial\Omega_k}\frac{\omega_1}{2}|\tilde{\mathbf{Q}}-\tilde{\mathbf{Q}}^{\parallel}|^2 + \frac{\omega_2}{2}(tr\tilde{\mathbf{Q}}^2-s_+^2)^2 \diff A,\ k = i\ or\ o,\nonumber
\end{equation}
on the inner and outer surfaces of the shell $\Omega$, $\partial\Omega_{i}$ and $\partial\Omega_{o}$,
where
\begin{gather}
\tilde{\mathbf{Q}} = \mathbf{Q}+\frac{s_+\mathbf{I}}{3},\ \tilde{\mathbf{Q}}^{\parallel} = \mathbf{P}\tilde{\mathbf{Q}}\mathbf{P},\ \mathbf{P} = \mathbf{I}-\mathbf{v}\otimes\mathbf{v},
\end{gather}
$\mathbf{v}$ is the unit normal vector, $\omega_1$ is the reduced anchoring strength that favours the tangential orientation of nematic director $\mathbf{n}$ i.e. prefers the leading eigenvector of $\mathbf{Q}$, labelled as the nematic director, to be in the plane of the spherical surface, and $\omega_2$ pushes $\mathbf{Q}$ towards the set of nematic or ordered bulk energy minimisers for a given $A < \frac{B^2}{24 C}$. Recall the definition of $s_+$ from \eqref{eq:s+}.

\begin{figure}[ht]
    \centering
    \includegraphics[width=0.5\textwidth]{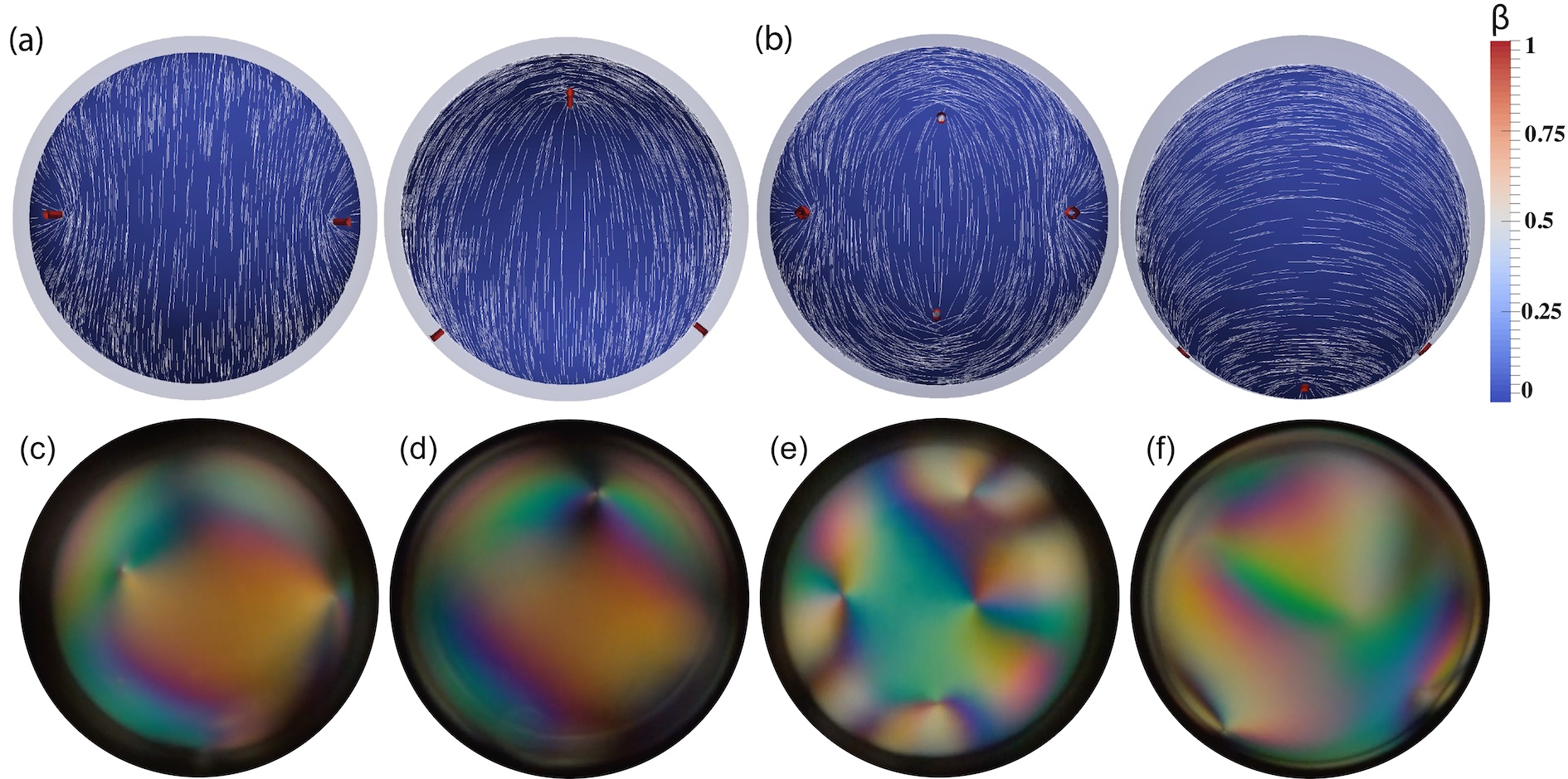}\\
    \caption{The numerical and corresponding experimental states with four +1/2 defects on the (a,c--d) symmetric (tetrahedral defect arrangement) and (b,e--f) asymmetric (all defects near the thinnest point) spherical shells. 
    In the numerical results (a-b), each left image is viewed from the thinnest side (bottom), and each right image is side viewed. The red pillars represent defect lines. The white lines represent the director $\mathbf{n}$ (eigenvector of $\mathbf{Q}$ with the largest eigenvalue). The coloring represents the biaxiality parameter $\beta = 1 - 6\frac{\big(tr\textbf{Q}^3\big)^2}{\big(tr\textbf{Q}^2\big)^3}$. $\beta= 0$ if $\mathbf{Q}$ is uniaxial. $\beta= 1$ corresponds to the case that biaxiality is maximal. The other parameters are $\rho = 0.8$, $t = -1.79$, $\xi_R = 1/50$, $\omega_1 = \omega_2 = 100$, $c=0$ for symmetric shell and $c=0.08$ for asymmetric shell. The POM images (c--f) show a symmetric (c--d) and an asymmetric (e--f) shell of nematic 8CB, each with diameter about 200~$\mu$m and average thickness about 10~$\mu$m. The polarizers are vertical and horizontal, respectively. The symmetric shell is imaged along the thickness gradient at $T=37.1^\circ$C with focus, respectively, near the bottom (c), clearly revealing two of the tetrahedrally arranged defects with the third hinted near the lower part of the image, and near the top (d), revealing the fourth defect. The asymmetric shell is viewed along the thickness gradient with focus near the bottom in (e), revealing all four defects ($T=40.5^\circ$C) and perpendicular to the thickness gradient with focus on the equator in (f), with two defects visible ($T=34.3^\circ$C). The photos are extracted from the Supporting Video.}

    \label{fig:TP}
\end{figure}

By minimizing the energy (refer to initial conditions and numerical methods in Appendix \ref{sec:b}), we obtain a state with four $+1/2$ defects (see Fig.~\ref{fig:TP}a and b). 
For a symmetric shell, i.e. $c = 0$ (Fig.~\ref{fig:TP}a), the state has the tetrahedral arrangement of four defects as reported in \cite{lopez2011frustrated,ishii2020structural}.
When the shell becomes more asymmetric, i.e. $c$, increases, the defect lines move to the thinnest part of the shell (Fig.~\ref{fig:TP}b).
We deduce that when $c\to1-\rho$, the four $+1/2$ defects  merge together to a $+2$ defect at the thinnest point of the shell, and we refer to this prototype state as the limiting state in subsequent discussions. 

Fig.~\ref{fig:TP}c--f shows polarizing optical microscopy (POM) photos of two shells of the liquid crystal material 8CB in the nematic phase, exhibiting the different defect configurations predicted from simulations. One shell is thick and almost symmetric, leading to a nearly tetrahedral distribution of the four defects, three of which are closer to the bottom (Fig.~\ref{fig:TP}c) and one of which is near the top (d). The other shell is somewhat thicker, leading to a clearly asymmetric shell thickness. Because the density of 8CB is nearly matched to that of the internal aqueous phase at a temperature just above the SmA-N transition, the plane containing the thinnest and thickest points becomes the horizontal plane \cite{Noh2021}, we can view the shell `from the side' at this temperature and thus distinguish the asymmetry, as shown in Fig.~\ref{fig:TP}f, corresponding to the simulation shown on the right in Fig.~\ref{fig:TP}b. We see one defect in focus and another can be distinguished albeit somewhat out of focus. When we heat further through the nematic phase, the density of 8CB gets lower than that of the inner aqueous phase, moving the internal droplet to the bottom, which becomes the thinnest point of the shell, while the top becomes the thickest point. When we focus at the bottom at such a temperature (Fig.~\ref{fig:TP}e) we see all four defects simultaneously, as they are collected near the thinnest point. The configuration is not identical to the ideal fully equilibrated configuration in the left simulation image in Fig.~\ref{fig:TP}b, but qualitatively the asymmetric shell fully resembles the corresponding simulated shell, with all four defects near the thinnest point.

\section{The $r$-axis/$\eta$-axis invariant reduced profiles}
\label{sec:OF}
Given our shell geometry: a shell with inner radius $R_i$, outer radius $R_o$, and the eccentricity i.e. the distance between the inner and outer spherical centers, $\delta$,  we use the spherical coordinates $(r,\theta,\phi)$ to describe the symmetric shell and bispherical polar coordinates $(\eta,\theta,\phi)$ (see Appendix \ref{sec:bispherical}) to describe the asymmetric spherical shell. $r$ is the radial spherical coordinate, $\eta$ is the corresponding radial bispherical coordinate, $\theta$ is 
the polar angular coordinate, and $\phi = \arctan(y/x)$ is the azimuthal angular coordinate.
The ranges of $r$, $\eta$, $\theta$ and $\phi$ are $r\in[R_i,R_o]$, $\eta\in[\eta_o,\eta_i]$, $\theta\in[0,\pi]$, $\phi\in[0,2\pi)$.

The eccentricity $\delta$ is an important parameter to measure the symmetry of a spherical shell.
When $\delta = 0$, the spherical shell is symmetric. When $\delta \to R_o-R_i$ , the spherical shell is super asymmetric with the thickness of the thinnest part tending to zero. 
The thinnest and thickest part of the shell correspond to $\theta = \pi$ and $\theta = 0$ in our spherical/bisperical coordinates respectively. 

In what follows, we assume that the average thickness of the shell (10~$\mu$m) is relatively small compared to the radius (100~$\mu$m), the state is $r$ or $\eta$-independent throughout the shell and the shell asymmetry only affects the defect locations. 

\subsection{The locations of defect lines}\label{sec:location}
According to the numerically computed tetrahedron states in Fig. \ref{fig:TP}(a-b), on both symmetric and asymmetric shells, the four lines on the shell surface connecting the defect, north pole and south pole divide the shell surface into four equal parts. So we assume that the four defects are located on $\phi = k\pi/2$, $k = 0,\cdots,3$.

\begin{figure}[ht]
    \centering
\includegraphics[width=0.5\textwidth]{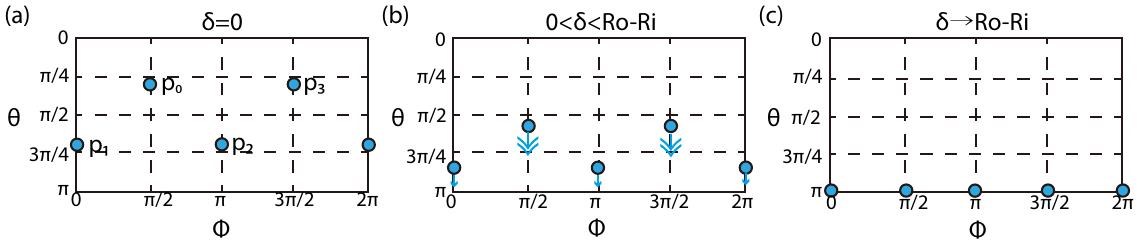}\\
    \caption{The schematic diagram of the location of defects for the state with four $+1/2$ defects in Fig. \ref{fig:TP}(a-b) when the eccentricity, (a) $\delta=0$, (b) $0<\delta<R_o-R_i$, and for the limiting state when (c) $\delta\to R_o-R_i$.}
    \label{fig:loc}
\end{figure}

For a symmetric shell with $\delta = 0$, the length of the defect lines are equal to $R_o-R_i$, and do not depend on $\theta$ (from the assumed $r$ or $\eta$-independence throughout the shell). Hence, the energy-minimizing locations of defects depend solely on the director field deformation around the defects. Due to the symmetric arrangement of defects on a symmetric shell in our numerical results in Fig. \ref{fig:TP}(a-b), we assume the location of defects on $\phi\theta$-plane are: $p_0 = (\pi/2,\pi/2-b)$, $p_1 = (0,\pi/2+b)$, $p_2 = (\pi,\pi/2+b)$ and $p_3 = (3\pi/2,\pi/2-b)$. Since the defects are evenly distributed, the geodesic distance on the spherical surface (the great circle distance) between $p_0$ and its three adjacent points $p_1,p_2,p_3$ are the same. So according to the calculation of the great circle distance in \cite{kells1940plane}, we have $b = arcsin(1/\sqrt{3})$. The location of defects are illustrated in Fig. \ref{fig:loc}(a).\\

For an asymmetric shell, i.e., $0 < \delta < R_o-R_i$, in the bispherical coordinate system $(\eta,\theta,\phi)$ (see  \eqref{eq:x}-\eqref{eq:z}),
the length of a $+1/2$ defect along the $\eta$ direction is
\begin{align}
\mathcal{L} &= \int_{\eta_o}^{\eta_i}\sqrt{\left(\frac{\partial x}{\partial \eta} \right)^2 + \left(\frac{\partial z}{\partial \eta}\right)^2} \diff\eta\nonumber\\
&= \int_{\eta_o}^{\eta_i}\frac{a\sqrt{\sin^2\theta\sinh^2\eta + (1-\cosh\eta\cos\theta)^2}}{(\cosh\eta-\cos\theta)^2} \diff\eta\nonumber\\
& = \int_{\eta_o}^{\eta_i}\frac{a}{(\cosh\eta-\cos\theta)} \diff\eta.\nonumber
\end{align}
 As $\theta$ increases to $\pi$, the length $\mathcal{L}$ decreases. The defect lines are energy unfavorable. Hence, in an asymmetric spherical shell, to reduce the length of the defect lines, the four defects tend to move to the thinner side of the shell as illustrated in Fig. \ref{fig:loc}(b).

For a super asymmetric shell, i.e., $\delta\to R_0-R_i$, 
the four defects merge and concentrate at $\theta = \pi$ as illustrated in Fig. \ref{fig:loc}(c).

\subsection{Profiles on $r = r^*$ or $\eta = \eta^*$}
For the numerical results in the LdG framework in Fig. \ref{fig:TP}(a-b), the states with four $+1/2$ defects are almost uniaxial and have almost constant orientational order far from the defects.
With the constraints of uniaxiality and constant orientational order, the LdG model can be reduced to the Oseen-Frank model \cite{landau2010beta}, which describes the nematic phase by a unit-vector field $\mathbf{n}$. Here, $\mathbf{n}$ models the uniaxial nematic director.
We can fix the location of defects according to the heuristic arguments presented in Section \ref{sec:location}, and assume $r$ or $\eta$-invariance on relatively thin shells so that the study of physically observable equilibria reduces to the following Oseen-Frank energy minimisation problem. On a surface $S$ with $\eta = \eta^*$ or $r = r^*$, the Oseen-Frank energy is given by \cite{frank1958liquid,oseen1933theory}:
\begin{equation}\label{eq:Frank} 
E(\mathbf{n}) = \int_{S}f_{el} \diff S = \frac{K}{2}\int_{S}(\nabla\cdot\mathbf{n})^2 + |\nabla\times\mathbf{n}|^2 \diff S,
\end{equation}
the first term describes splay deformations, and the second term describes the twist and bend deformations of $\mathbf{n}$. $K$ is an elastic constant.

Due to the experimentally imposed tangential boundary condition on the outer and inner surfaces, we assume that the nematic directors are tangential to an intermediate surface, $r = r^*$ or $\eta = \eta^*$ as well i.e. $\mathbf{n}$ is of the form:
\begin{equation}\label{eq:alpha}
\mathbf{n} = \sin\alpha(\theta,\phi) \mathbf{e}_{\theta} + \cos\alpha(\theta,\phi)\mathbf{e}_{\phi},
\end{equation} where $\mathbf{e}_\theta$ and $\mathbf{e}_\phi$ are an orthonormal basis on a spherical surface.
Subsequently, the elastic energy density of \eqref{eq:Frank} is 
\begin{align}\label{eq:elastic_density}
 &f_{el} = \frac{K}{2}((\nabla\cdot\mathbf{n})^2 + |\nabla\times\mathbf{n}|^2)r^2\sin \theta\nonumber\\
 &= \frac{K}{2}\left(\frac{(\cos\theta-\partial_{\phi}\alpha)^2}{\sin\theta} + (\partial_{\theta}\alpha)^2\sin\theta\right).
\end{align}
The energy minimisers are solutions of the corresponding Euler-Lagrange equation given by
\begin{equation}\label{eq:euler_lagrange}
 \partial^2_{\phi} \alpha/\sin\theta + \partial^2_{\theta}\alpha \sin\theta = 0.
\end{equation}

\begin{figure}[ht]
    \centering
\includegraphics[width=0.2\textwidth]{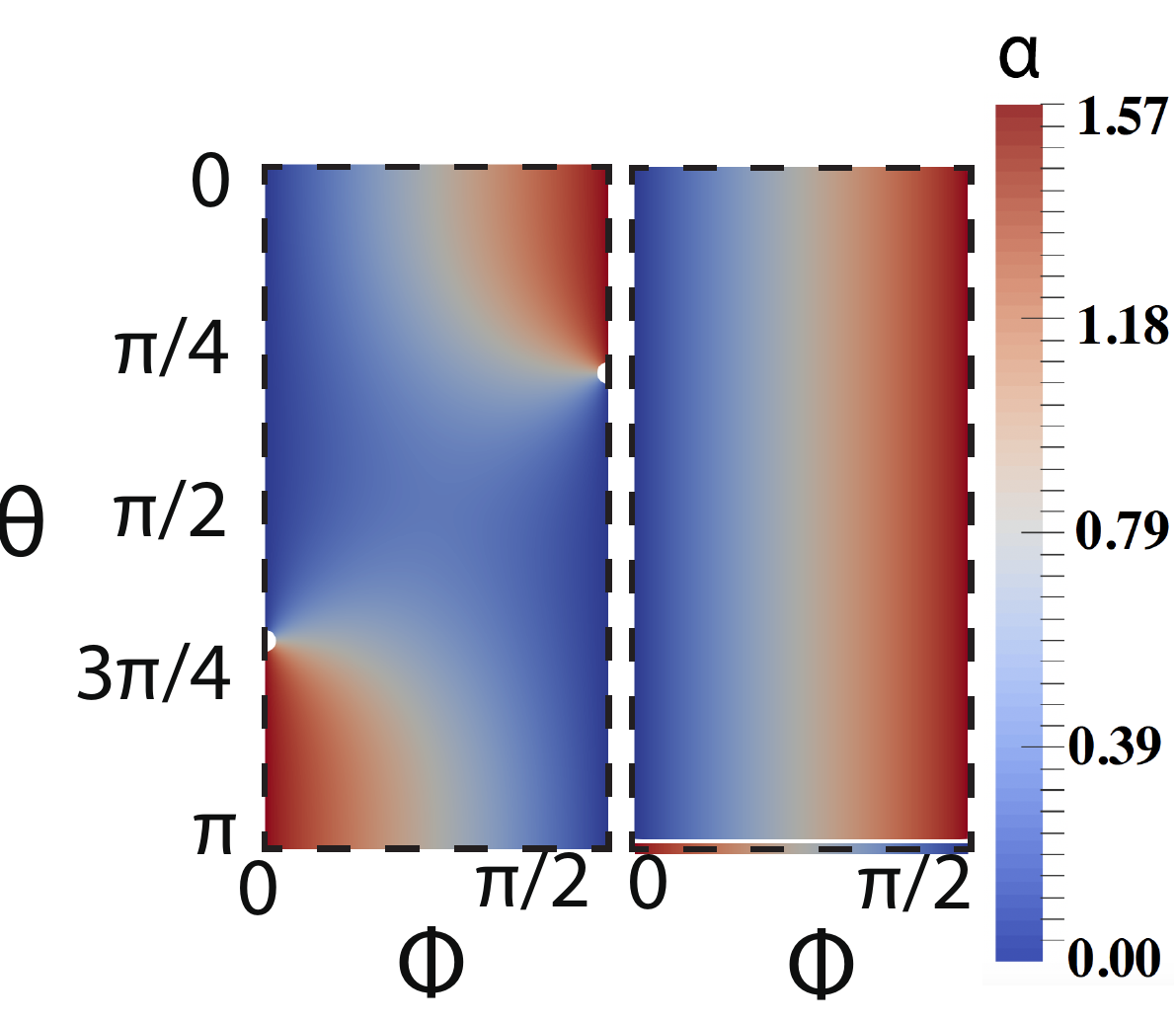}\\
    \caption{The plot of solution $\alpha$ for symmetric shell ($\delta=0$) and asymmetric shell ($\delta\to R_0-R_i$). Left: the solution of \eqref{eq:euler_lagrange} with boundary conditions \eqref{eq:32}-\eqref{eq:39} on the domain $D_{\phi\theta} = [0,\pi/2]\times[0,\pi]\backslash \{D((0,\theta_1),0.05)\cup D((\pi/2,\theta_2),0.05)\}$.
    Right: $\alpha = \phi$ in \eqref{eq:alpha_phi}, the solution of \eqref{eq:euler_lagrange}, with boundary conditions \eqref{eq:40}-\eqref{eq:43} on the domain $D_{\phi\theta}=[0,\pi/2]\times[0,\pi-0.05]$. The black dashed line outlines the area of $[0,\pi/2]\times[0,\pi]$.}
    \label{fig:solution}
\end{figure}

The $\theta = 0$ and $\theta = \pi$ coordinates are two points on a spherical surface but they are two lines in the $\phi\theta$-plane. How do we define the appropriate boundary conditions for these antipodal points? 
Substituting $\theta = 0$ and $\pi$ into \eqref{eq:elastic_density} and ensuring that the energy density doesn't diverge,
we impose the boundary condition
\begin{gather}
\partial_{\phi}\alpha = 1,\ \textrm{on}\ 
 \theta = 0,\ \partial_{\phi}\alpha = -1,\ \textrm{on}\ 
  \theta = \pi.
\end{gather}
Due to the symmetry of the tetrahedron state in Fig. \ref{fig:loc}(a), in the following, we reduce the domain of $(\phi,\theta)$ to a quarter of the domain i.e. $[0,\pi/2]\times[0,\pi]$.
In the symmetric limit $\delta=0$, following the discussion in Section \ref{sec:location} and Fig \ref{fig:loc}(a), we assume that the defects are located at the points $p_1 = (0,\theta_1)$ with $\theta_1 = \pi/2+arcsin(1/\sqrt{3})$ and $p_0 = (\pi/2,\theta_2)$ with $\theta_2 = \pi/2-arcsin(1/\sqrt{3})$, in the computational domain. 
After removing a small neighbourhood of the defect with core size $\epsilon$, the domain is $D_{\phi,\theta}=[0,\pi/2]\times[0,\pi]\backslash \{D((0,\theta_1),\epsilon)\cup D((\pi/2,\theta_2),\epsilon)\}$.
According to the configuration of the tetrahedron state in Fig. \ref{fig:TP}, we assume that $\alpha$ is continuous on the boundary of the domain far from defects, the director is vertical or horizontal (on the $\phi\theta$-plane) on $\phi = 0$ and $\phi=\pi/2$, $\alpha$ jumps $\pm \pi/2$ when crossing the $+1/2$ defects on $\phi = 0$ and $\phi=\pi/2$.
The boundary conditions on $\partial D_{\phi,\theta}$ are explicitly written in the Appendix \ref{sec:bc}.
The numerical solution of \eqref{eq:euler_lagrange} with the appropriate boundary conditions in \eqref{eq:32}-\eqref{eq:39}, corresponding to the tetrahedral defect arrangement on a symmetric shell, is shown in Fig. \ref{fig:solution} on the left. We can obtain $\alpha$ throughout the whole domain by means of reflection symmetry, and the corresponding director $\mathbf{n}$ through \eqref{eq:alpha}.
In this reduced study, we get a nonlinear solution $\alpha$ in Fig. \ref{fig:solution} rather than a simple $\alpha = -\phi/2$ corresponding to a single $+1/2$ defect as in Section \ref{sec:clearing}. This nonlinear solution captures the collective effects or interactions of the four defects in the tetrahedron state on a symmetric shell and provides a relatively simple method for capturing the qualitative features of the computationally demanding numerical solutions in Fig.~\ref{fig:TP}. Notably, we only solve a boundary-value problem for $\alpha$ on a truncated rectangular domain as opposed to a system of five nonlinear partial differential equations on a three-dimensional shell domain, subject to weak anchoring conditions in Fig.~\ref{fig:TP}. 

Again, following the discussion in Section \ref{sec:location} and Fig \ref{fig:loc}(c),
in the super asymmetric limit $\delta\to R_o-R_i$,  the defects concentrate near $\theta = \pi$. We cut the domain near $\theta = \pi$ to study the domain $D_{\phi\theta}=[0,\pi/2]\times[0,\pi-\epsilon]$ with the boundary conditions
\begin{align}
 &\alpha = 0,\ \text{if}\ \phi = 0,\ \alpha = \pi/2,\ \text{if}\ \phi = \pi/2,\label{eq:40}\\
 &\alpha = \phi,\ \text{if } \theta = 0\ \text{and}\ \theta = \pi-\epsilon.\label{eq:43}
 \end{align}
These are the limiting version of the boundary conditions in \eqref{eq:32}-\eqref{eq:39} as $\theta_1\to\pi$ and $\theta_2\to\pi$ respectively.
 
The corresponding analytic solution in the limit $\delta\to R_o-R_i$ is
\begin{equation}\label{eq:alpha_phi}
\alpha = \phi.
\end{equation}
The corresponding numerical solution is shown in Fig. \ref{fig:solution} on the right.
There is a $+2$ point defect at $\theta = \pi$, which is the south pole of the shell. This analysis is complementary to our full numerical results. The full numerical results in Fig. \ref{fig:TP}(a-b) show that, as the shell gets increasingly asymmetric (or thicker), the four $+1/2$ defects get closer to the thinnest point. However, we cannot use a 
numerical method to simulate the limiting case with $\delta = R_o-R_i$, and get a real $+2$ defect on a shell, since the thickness of the thinnest part is zero for this limiting situation.



\section{Experimental Analysis and Modelling of the Clearing Transition}\label{sec:clearing}
\begin{figure}[t!]
    \centering
\includegraphics[width=\columnwidth]{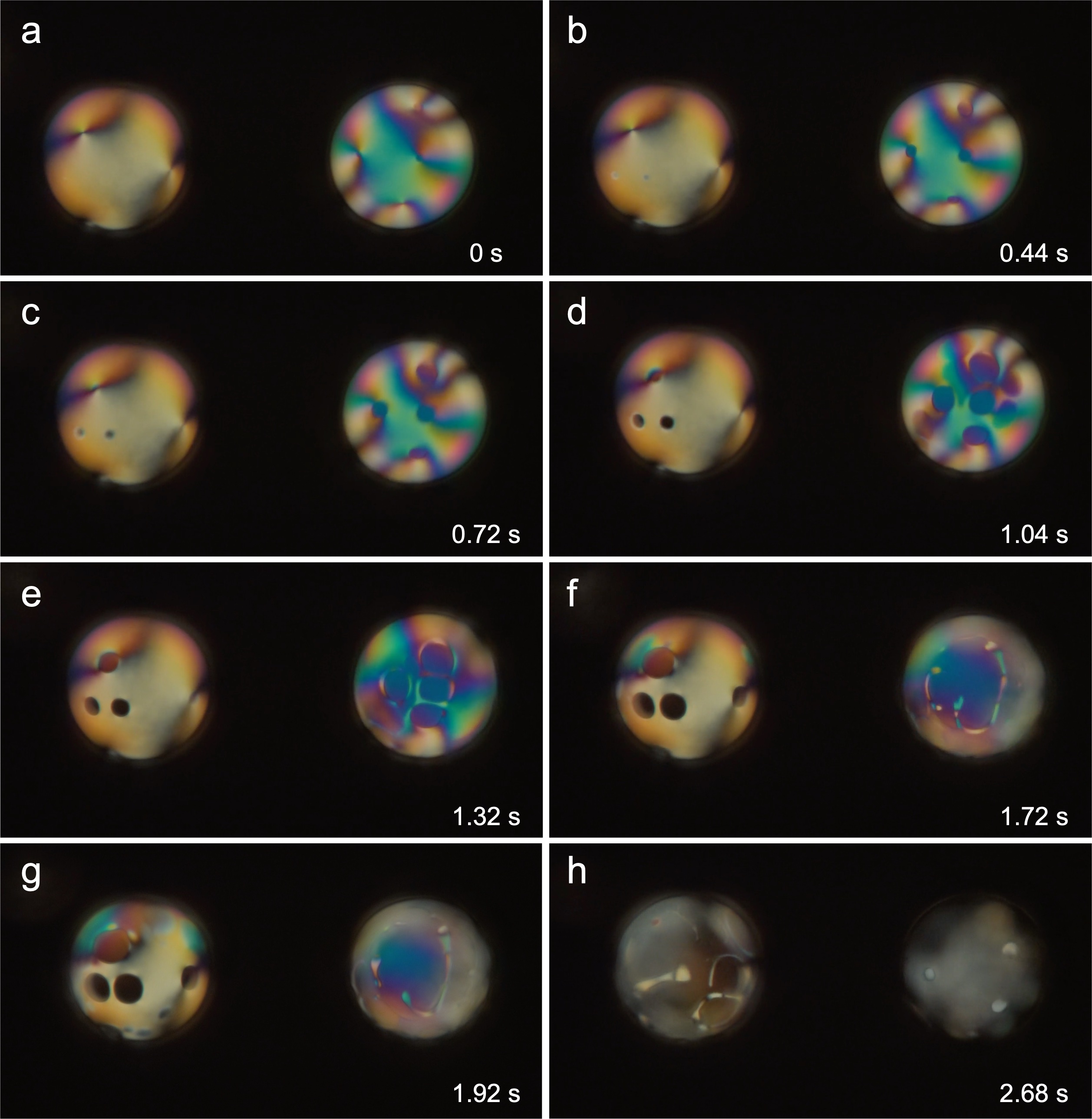}\\
    \caption{The same two shells (symmetric on the left, asymmetric on the right) as in Fig.~\ref{fig:TP}c--f, as they go through the nematic--isotropic transition. The focus is near the shell bottoms and the sample chamber is heated at a constant rate of 1~K/min., the hot stage reading being $40.5^\circ$C in (a)--(g) while in (h) it is $40.6^\circ$C. Time stamps at bottom right of each frame refer to the time after the first sign of the transition is detected, in the right shell. The photos are still frames from the Supporting Video.}
    \label{fig:nucleation}
\end{figure}

We now study the clearing transition in the shells, experimentally and numerically. Considering the experimental study first, Fig.~\ref{fig:nucleation} shows the same two shells as in Fig.~\ref{fig:TP}(c-f) as they go through the nematic--isotropic transition during heating at 1~K/min. Each frame of this figure shows both shells at the same time and temperature, revealing that the transition starts slightly earlier in the asymmetric shell, which is on the right in each image. In (a), which we consider to be time 0~s, the transition can for the first time be detected in the asymmetric shell, at three isotropic nuclei at defects. The symmetric shell is still fully nematic here, its first isotropic nuclei appear outside defects, in (b). Its first isotropic nucleus in a defect is seen in (c). In (d) we see the isotropic phase nucleate on the thick side of the asymmetric shell (out of focus) and in (e) the symmetric shell nucleates the transition in its second defect visible in this picture. In (f) the isotropic domains have started merging in the asymmetric shell and a color change in the upper left of the symmetric shell reveals that the transition has started on the side out of focus. As more and more of the shell turns isotropic, the nematic boundaries that separated the first nuclei in the asymmetric shell collect into isolated islands (g) which end up being the last remaining points of nematic state before the entire shell goes isotropic (h).

We conclude from this experiment that (1) the clearing transition starts at slightly lower temperature in the asymmetric shell for which all four defects are collected near the thinnest point, and (2) the four defects function as the very first nucleation points for the transition in the asymmetric shell. In contrast, (3) when the shell is nearly symmetric, the first isotropic nuclei can appear anywhere, also outside defects, but each defect still acts as a nucleating point of the transition early in the process, while the shell is mainly in the nematic phase.

\begin{figure}[t!]
    \centering
\includegraphics[width=\columnwidth]{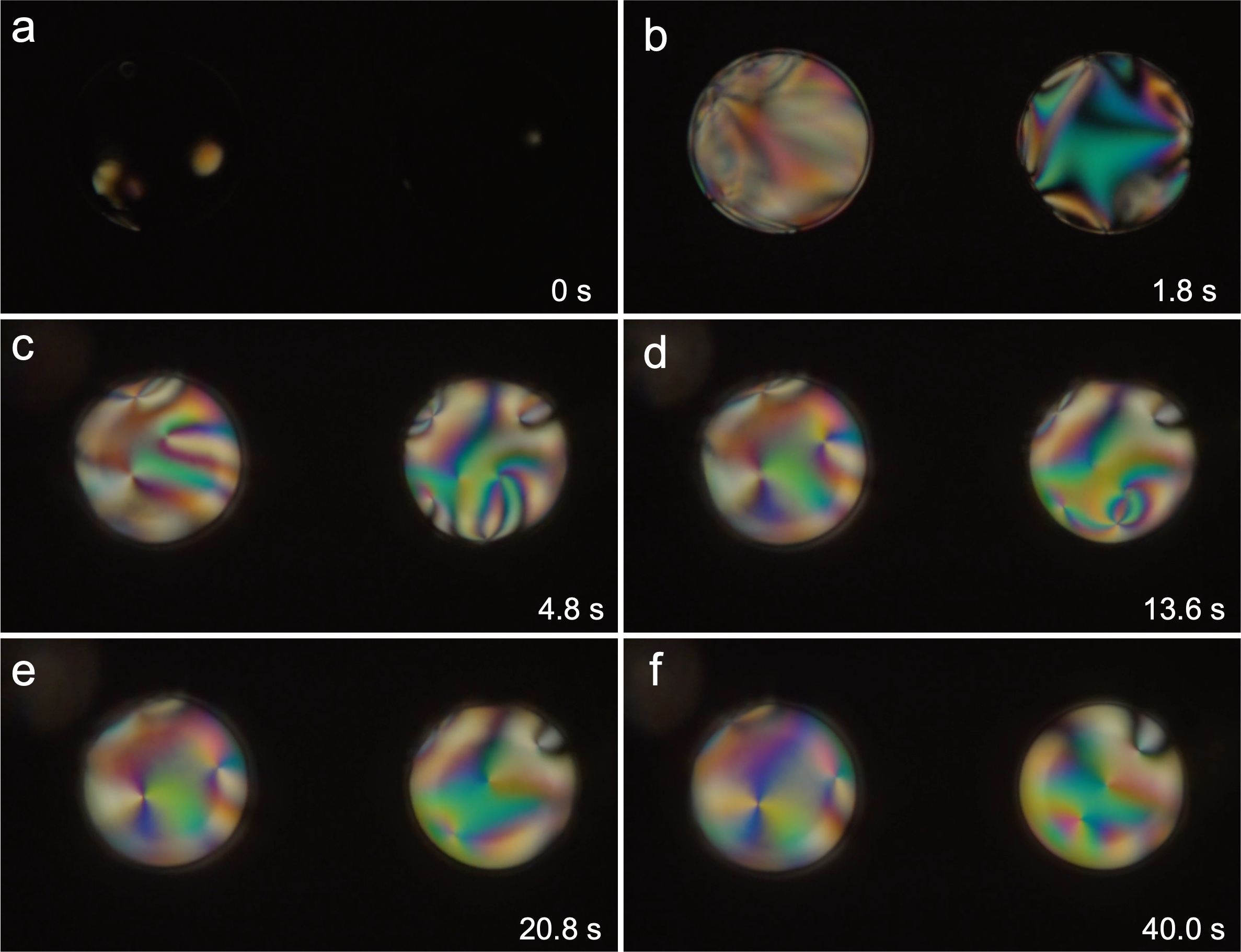}\\
    \caption{The same two shells being cooled through the isotropic--nematic transition. The focus is near the shell bottoms and the sample chamber is cooled at a constant rate of 1~K/min., the hot stage reading being $40.5^\circ$C in (a)--(b), $40.4^\circ$C in (c), $40.3^\circ$C in (d), $40.2^\circ$C in (e) and $39.9^\circ$C in (f). Time stamps at bottom right of each frame refer to the time after the first sign of the transition is detected in the right shell, slightly after the transition has started in the left shell. The photos are still frames from the Supporting Video.}
    \label{fig:coolingtransition}
\end{figure}

Once the shells are entirely isotropic, the sample is cooled at -1~K/min. and we follow the transition back into the nematic phase in Fig.~\ref{fig:coolingtransition}. Corresponding to the left shell retaining nematic phase to the highest temperature in Fig.~\ref{fig:nucleation}, it is also the first shell to nucleate the nematic phase on cooling; in Fig.~\ref{fig:coolingtransition}a the first nuclei in this phase have already grown to respectable size while we see the very first nematic nuclei in the right shell. We believe that the subtle difference between the two shells, consistent between the heating and cooling experiments, is due to the slightly greater average thickness in the right asymmetric shell, meaning that it contains slightly more LC, thus exhibiting a slightly greater latent heat for the transition for the overall shell.

As the shells have turned entirely nematic (Fig.~\ref{fig:coolingtransition}b--c) we first note that neither shell shows any memory of the defect configuration prevailing prior to the nematic--isotropic transition on heating. In fact, both shells clearly exhibit more than the topologically required defects, the symmetric shell showing one integer defect (strength $\pm 1$, four dark brushes) and two half-integer defects (strength $\pm 1/2$, two brushes) and the asymmetric shell showing one integer defect and five half-integer defects in focus in Fig.~\ref{fig:coolingtransition}c, additional defects most likely being present on the shell side out of focus. Since the total topological defect strength must still be +2 over the entire shell surface, we can conclude that some of the defects are negative-signed, as a result of multiple independently nucleated nematic phase domains merging upon cooling. In (d) we can see two integer defects (four brushes each) approaching each other in the right image, clearly attracting due to their opposite signs. In (e) they have annihilated and we see only three half-integer defects (most likely +1/2) moving towards the thinnest side, while the +1 defect is moving towards the opposite side, being barely visible. Most likely it is attracted to a half-integer defect of opposite sign on the side out of focus. In (f) it can no longer be seen and one may suspect that it has merged with the attractor defect, leaving only the final stable +1/2 defect that will eventually move back to the thinnest part of the shell.

The defects on the symmetric shell (on the left) also move during this process, but the progress towards the energy-minimizing tetrahedral configuration is slower than that for the corresponding asymmetric shell on the right. 
The asymmetry thus appears to drive a change toward an energy-minimizing defect configuration faster than in the symmetric shell. However,  the change is not particularly fast in the asymmetric shell; 40~s after the cooling transition process is initiated, the director field has not yet reached the stable configuration with all four defects concentrated near the thinnest point. 

We now turn to the theoretical analysis of the heating nematic--isotropic (NI) transition for the two shells. A relatively simple model was developed in \cite{mottram2000defect,mottram1997disclination} to track the critical temperature of the NI transition and the growth of isotropic phase, as a function of temperature, on the plane, assuming that the NI interface is a circle. We adapt this model to a spherical surface or a radially-invariant shell geometry, as shown below. 

\begin{figure}[ht]
    \centering
\includegraphics[width=0.2\textwidth]{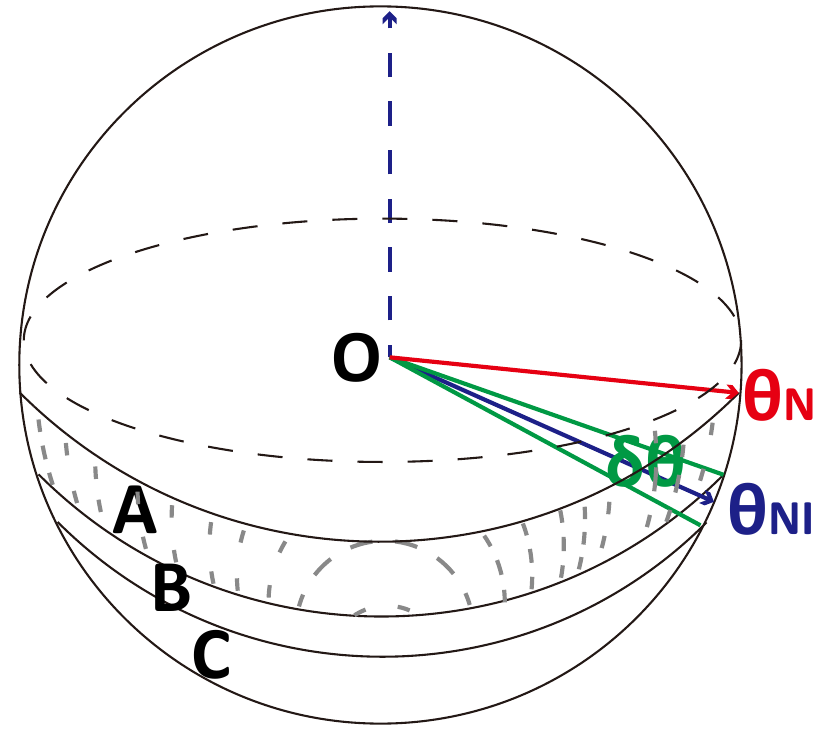}\\
    \caption{The simplified picture of a melting shell:
region C, the inner, isotropic, core of the defect $\theta_{NI}+\delta\theta/2<\theta<\pi$; region B, 
the nematic–isotropic (NI) interfacial region $\theta_{NI}-\delta\theta/2<\theta<\theta_{NI}+\delta\theta/2$; region A, the distorted nematic region $\theta_N<\theta<\theta_{NI}-\delta\theta/2$.}
    \label{fig:ABC_domain}
\end{figure}

We consider the region $\theta_N\leq\theta\leq\pi$, assuming only one interior defect (+2 or +1/2) at the south polar point ($\theta = \pi$). The $+2$ defect case corresponds to an asymmetric shell whereas the $+1/2$ defect case corresponds to a symmetric shell (with a regular tetrahedral arrangement of four $+1/2$ defects). Following the experimental video, $\theta_N>\pi/2$ and we take $\theta_N$ to be constant throughout the manuscript. For fixed $r = r^*$ or $\eta = \eta^*$, the free energy on $(\phi,\theta)\in[0,2\pi)\times[0,\pi]$ can be written as
\begin{align}
F &= \int_0^{2\pi}\int_{\theta_{NI}+\delta\theta/2}^{\pi} f_I\sin\theta \bar{r}^2 \diff\theta \diff\phi + \int_0^{2\pi}\sigma_{NI}\sin\theta_{NI}\bar{r}\diff\phi \nonumber\\
&+ \int_0^{2\pi}\int_{\theta_{N}}^{\theta_{NI}-\delta\theta/2} f_N\sin\theta \bar{r}^2 \diff\theta \diff\phi \nonumber\\
&+ \frac{K}{2}\int_0^{2\pi}\int_{\theta_N}^{\theta_{NI}-\delta\theta/2}\frac{(\cos\theta-\partial_{\phi}\alpha)^2}{\sin\theta} + (\partial_{\theta}\alpha)^2\sin\theta \diff\theta \diff\phi.
\end{align}
where $\theta_{NI}$ is the location of NI interface and $\delta\theta$ is the width of the NI interface (see Fig \ref{fig:ABC_domain}), $f_I$ and $f_N$ are the entropic free energy per unit volume of the isotropic and nematic phases respectively, $\sigma_{NI}$ is the nematic-isotropic surface free energy per unit area, $K$ is the Oseen-Frank elastic constant (analogous to the elastic constant $K$ in \eqref{eq:Frank}), $\alpha$ is the angle of director defined in \eqref{eq:alpha}, $\bar{r}$ is the radius of the studied spherical surface with $r = r^*$ or $\eta = \eta^*$.

We define $T_{NI}$ as the transition temperature for which the bulk energy of nematic and isotropic phases are the same, $f_I(T_{NI})=f_N(T_{NI})$. To match with the experimental data, we use $T_{NI}=40.5^\circ$C for 8CB \cite{weiss2000orientation}.
We simplify the quantities $f_I$ and $f_N$ close to $T_{NI}$ using a Landau expansion for each phase,
\begin{align}
f_{I}(T) &= f_I(T_{NI})-S_I(T-T_{NI})+O(T-T_{NI})^2,\nonumber\\
f_{N}(T) &= f_N(T_{NI})-S_N(T-T_{NI})+O(T-T_{NI})^2,\nonumber
\end{align}
where $S_N$ and $S_I$ are the entropy per unit volume per kelvin of nematic and isotropic phases respectively.

With $f_I(T_{NI}) = f_N(T_{NI})$, we obtain
\begin{equation}
f_I-f_N = -(S_I-S_N)(T-T_{NI}) + O(T-T_{NI})^2.
\end{equation}
Neglecting higher order terms of $T-T_{NI}$ and $\delta\theta$ (to
within an irrelevant constant)
\begin{align}
F(\theta_{NI}) &= -2\pi\Delta S(T-T_{NI})\bar{r}^2\cos\theta_{NI} \label{eq:F_NI}\\
&+ 2\pi\sigma_{NI}\bar{r}\sin\theta_{NI} \\
&+ \frac{K\pi}{2}((d-1)^2ln(1-\cos\theta_{NI})+2\cos\theta_{NI}\label{eq:K1}\\
&-(d+1)^2ln(1+\cos\theta_{NI})).\label{eq:K2}
\end{align}
with $\Delta S = S_I-S_N>0$ (the entropy of isotropic phase is always higher than nematic phase), and $d = \partial_{\phi}\alpha$, ($\alpha = \phi$ and $d = 1$ for $+2$ point defect,  $\alpha = -\phi/2$ and $d = -1/2$ for $+1/2$ point defect at the south pole, the winding number of $\alpha = d\phi$ near the south pole is calculated in Appendix \ref{sec:d}).
Since the region under consideration is $\theta_N\leq\theta\leq\pi$, $\theta_N>\pi/2$ and $\theta_{NI}>\theta_N$, we have $\pi/2\leq\theta_{NI}\leq\pi$. Looking at the first term, if $\theta_{NI}$ decreases so that the size of the isotropic region increases, then the energy decreases for $T>T_{NI}$ and increases for  $T<T_{NI}$, consistent with the fact that the isotropic phase is energetically preferred for temperatures $T> T_{NI}$ and energetically unfavourable for $T< T_{NI}$. 
The second term implies that the presence of a nematic-isotropic interface leads to an increase in the free energy of the system. The third term favors a large isotropic area. For a fixed $\theta_N$, as $\theta_{NI}$ decreases, the elastic energy contained in the nematic region modelled by, $\theta_N < \theta < \theta_{NI}$, decreases.

We set the parameters to be:  the average radius of shell  $\bar{r} = 100$~$\mu$m = $10^{-4}$~m, entropy difference between isotropic and nematic phase $\Delta S = 2.91$~J$\cdot$mol$^{-1}\cdot$ K$^{-1}$/291.4 g$\cdot$ mol$^{-1}\times$ 0.985 g$\cdot$ cm$^{-3} \approx 10^4$~J$\cdot$m$^{-3}\cdot$~K$^{-1}$ (NI entropy difference 2.91~J$\cdot$ mol$^{-1}\cdot$~K$^{-1}$ in \cite{sharma2010non} Table 2, molar mass of 8CB 291.4 g$\cdot$ mol$^{-1}$, density $\rho = 0.985$~g$\cdot$ cm$^{-3}$ in \cite{sharma2010non} Table 3), the tension $\sigma_{NI} = 10^{-7}$~J$\cdot$ m$^{-2}$ in \cite{kim2013morphogenesis,faetti1984nematic}, the elastic constant $K = 2\cdot 10^{-12}$~J$\cdot$ m$^{-1}$ \cite{faetti1984nematic}.
The physically relevant solutions can be found by analytically minimizing
$F(\theta_{NI})$ in \eqref{eq:F_NI}-\eqref{eq:K2}. 

\begin{figure}[ht]
    \centering
\includegraphics[width=0.5\textwidth]{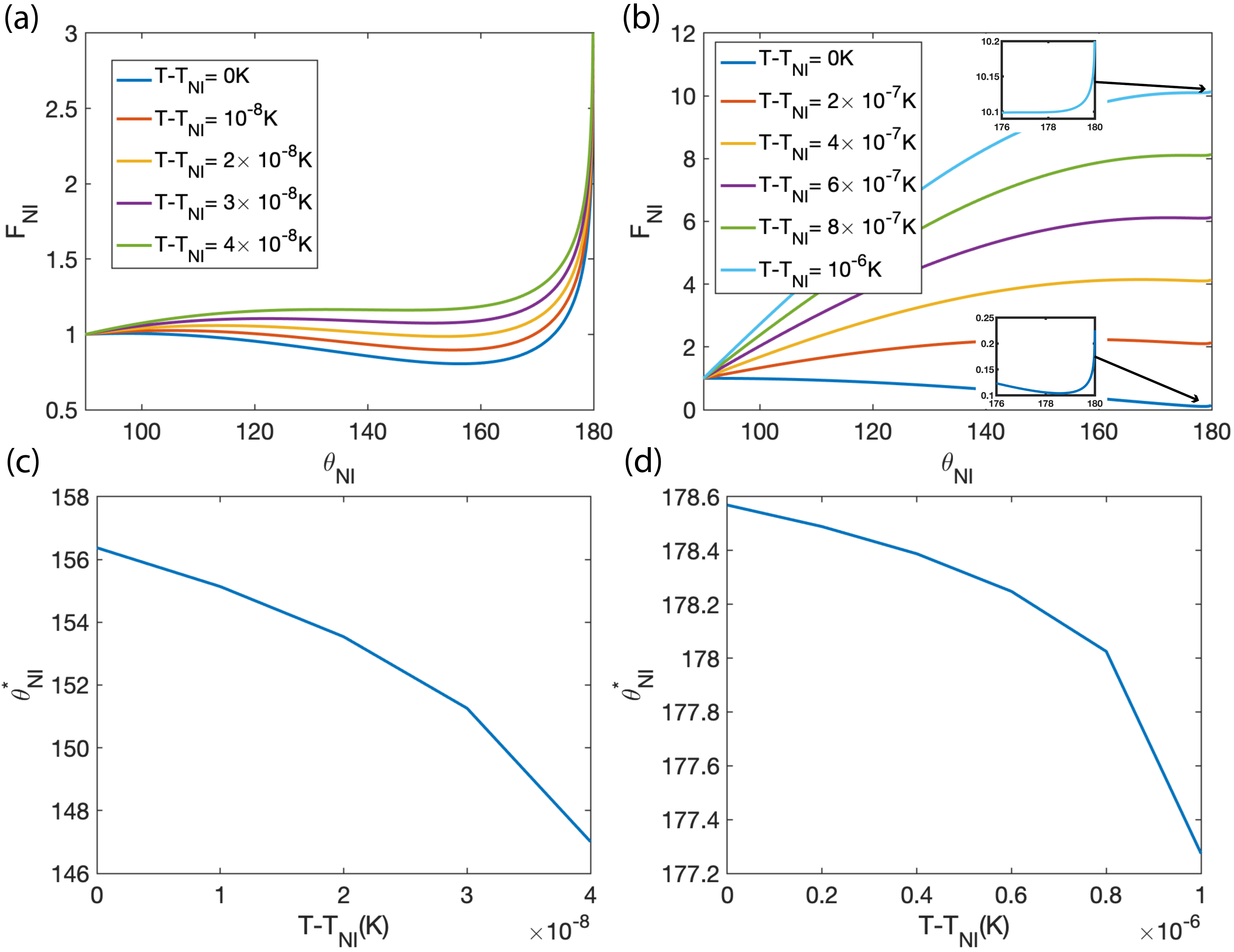}\\
    \caption{
    The plots of $F_{NI} = \frac{10^{11}}{2\pi}F(\theta_{NI})$ in \eqref{eq:F_NI} for (a) $+2$ defect and (b) $+1/2$ defect, and the local minimizer $\theta_{NI}^*$ of $F_{NI}$ versus $T-T_{NI}$ for (c) $+2$ defect and (d) $+1/2$ defect. As $T-T_{NI}$ increases, the minimiser $\theta_{NI}^*$ decreases, i.e. the area of isotropic increases.}
    \label{fig:NI}
\end{figure}

Before heating, the shell is in nematic phase, i.e., the NI interface is at $\theta_{NI} = \pi$.
As temperature $T$ increases, the NI interface moves to $\theta_{NI} = \theta_{NI}^*$, where $\theta_{NI}^*$ is the local minimiser of $F(\theta_{NI})$. For the asymmetric shell ($+2$ defect), as $T-T_{NI}$ increases from zero to $4\cdot 10^{-8}$ K, $\theta_{NI}^*$ decreases from around $156^{\circ}$ to around $147^{\circ}$ in Fig. \ref{fig:NI}(a) and (c). For the symmetric shell ($+1/2$ defect), as $T-T_{NI}$ increases from zero to $10^{-6}$ K, $\theta_{NI}^*$ decreases from around $179^{\circ}$ to around $177^{\circ}$ in Fig. \ref{fig:NI}(b) and (d). In both cases, when $T-T_{NI}$ increases further, the local minimizer $\theta_{NI}^*$ does not exist, and the NI interface jumps to the edge of the studied region $\theta_{NI} = \theta_{N}$, i.e., the shell is in isotropic phase. This critical temperature is the so-called clearing temperature, above which we speculate that the shell is totally isotropic.

We deduce from Fig. \ref{fig:NI} that the clearing temperature of an asymmetric shell ($+2$ defect) is $T_c = T_{NI}+O(10^{-8})$~K and the clearing temperature for a symmetric shell ($+1/2$ defect) is $T_c = T_{NI}+O(10^{-7})$~K (we cannot find solutions, $\theta_{NI}^*$, for temperatures $T - T_{NI} = 1.1\times 10^{-6}$ ~K and hence, we deduce that the clearing temperature is of order $T_{NI} + O(10^{-7}) $~K on the symmetric shell). Below the clearing temperature, the maximum isotropic area for the asymmetric shell ($+2$ defect) is between $\theta\approx 147^{\circ}$ and south pole, and the maximum isotropic area for the symmetric shell ($+1/2$ defect) is between $\theta\approx 177^{\circ}$ and south pole. 

In particular, when there is no defect at the south polar point, i.e., $\alpha = -\phi$ and $d =-1$, the elastic energy density in \eqref{eq:K1}-\eqref{eq:K2} near $\theta_{NI} = \pi$ is much lower and changes more gently with $\theta_{NI}$ than the elastic energy density for $+2$ and $+1/2$ defects, with $d = 1$ and $d = -1/2$ respectively. As $F'(\pi) = - 2\pi\sigma_{NI}\bar{r} <0$, $\theta_{NI} = \pi$ is always a minimizer of $F(\theta_{NI})$ 
with $d = -1$, which means the NI transition cannot occur unless a perturbation creates or nucleates a small isotropic domain around $\theta_{NI} = \pi$ that triggers the NI transition.

These numerical and analytic results agree with the experimental results in the sense that the clearing temperature is lower for an asymmetric shell, i.e. the clearing transition starts at slightly lower temperatures in the asymmetric shell,  the maximum isotropic area for the asymmetric shell ($+2$ defect) is larger than that for the symmetric shell ($+1/2$ defect) below the respective clearing temperatures (see Fig. \ref{fig:nucleation}), and defects function as nucleation points for the transition due to concentration effects of the elastic energy. We have not modelled the reverse cooling transitions in this section.


\section{Conclusions and Discussion}
\label{sec:conclusions}
We have demonstrated, experimentally and theoretically, that topological defects in nematic shells act as nucleation points for the transition to isotropic phase upon heating, as can be expected given that each topological defect acts like a local region of isotropic phase even at temperatures where the bulk nematic phase is absolutely stable in the absence of confinement effects. When all defects are collected close to each other, as near the thinnest point of an asymmetric shell, the strong deformation of the director field around the defects additionally leads to a local reduction of the effective clearing temperature in this region, ensuring that the defects become the primary nuclei of the transition as the shell is heated into the isotropic phase. The effect is very small and hence nuclei, away from defects, are soon seen as the asymmetric shell is continuously heated in experiments. For a symmetric shell the local director field deformation is never strong enough to induce an experimentally detectable lowering of the effective clearing temperature, explaining why the isotropic phase nucleates outside topological defects in a symmetric shell. We propose a simple mathematical model, based on the arguments in \cite{mottram2000defect, mottram2014introduction}, to explain the heating transitions and whilst our model is not in perfect agreement with experiments, it captures the fact that the clearing transition proceeds more quickly in the asymmetric shell and at lower temperatures in asymmetric shells, compared to the symmetric counterparts i.e. the asymmetric shell will relax to an isotropic shell before the symmetric shell, during the heating transition. The faster relaxation to the isotropic phase is facilitated by the elastic energy concentration near the four defects around the south pole of an asymmetric shell. We propose a simple free energy on a spherical surface with entropic contributions, a NI-interface energy and an elastic distortion energy. The elastic distortion energy can distinguish between an asymmetric and symmetric shell, captured by the parameter $d$ in \eqref{eq:K1}-\eqref{eq:K2}, or the topological strength of the defect at the south pole. The elastic distortion energy favours an isotropic phase, with the effect being more pronounced for an asymmetric shell (with $d=+1$) compared to a symmetric shell (with $d=-1/2$) in \eqref{eq:K1}-\eqref{eq:K2}, and is one of the primary drivers of the relatively fast relaxation process in an asymmetric shell, within the remit of our simple model. Our modelling approaches are limited in numerous ways. For example, in Section~\ref{sec:LDG}, it is perfectly possible that there are multiple LdG energy minimisers on asymmetric and symmetric shells, with different defect configurations, with tangential anchoring and we have simply found one of the energy minimisers. The simple model for the clearing transition in Section~\ref{sec:clearing} neglects the effects of the shape of the NI interface or the actual dynamics of the NI interface, which could play a crucial role in the defect dynamics during heating and cooling transitions. However, these simple models do capture the essential experimental details. 

Although the impact of the defects is small in terms of shift in practical transition temperature, the fact that the transition always nucleates in defects, as primary nuclei when the defects are co-localized, can have practical implications in contexts where the NI transition in shells is used for applications. Examples are liquid crystal elastomer shell actuators \cite{fleischmann2012one,Sharma2021}, the strong shape morphing of which is driven by the nematic--isotropic phase transition, as well as liquid crystal-based sensors where the analyte induces this transition \cite{Ramou2023}, or where topological defects otherwise play a critical role for the detection \cite{Carlton2013}.

An intriguing aspect that we did not address in this paper is that, as the isotropic phase grows from a nucleus it may be considered to form a topological hole in the shell, thus connecting the in- and outsides with each other via the nematic-isotropic boundary and suddenly transforming the remaining nematic to a single-interface volume. However, the isotropic phase on the other side of this new interface is not constant, but changes from the aqueous isotropic phase on the shell in- and outsides to the isotropic phase of the LC material in the growing isotropic regime. These two different bounding phases are likely to impose different boundary conditions and/or different anchoring strengths, complicating the application of the Poincaré-Hopf theorem on the transitional single-interface nematic state. Alternatively, the isotropic phase might nucleate from the out- or the inside without reaching all the way through the shell, thus maintaining the overall topology as long as the nematic phase forms a continuous spherical surface at least at some plane. It is a stimulating challenge for future investigations to probe the nature of the phase transition with such resolution that the actual scenario can be identified.

\begin{acknowledgments}
The raw video footage used for the experimental part was obtained by Dr. JungHyun Noh in the context of a different research project while she was a PhD candidate under the supervision of JL. AM is supported by the University of Strathclyde New Professors Fund, the Humboldt Foundation and a Leverhulme Research Project Grant RPG-2021-401. YH is supported by the Sir David Anderson Bequest Award at University of Strathclyde and a Leverhulme Project Research Grant RPG-2021-401.
YH also thanks JL's group members Yong Geng, Yansong Zhang, and Xu Ma for interesting discussions.

\end{acknowledgments}

\appendix
\section{Bispherical polar coordinate}\label{sec:bispherical}
We use the geometric conversions
between Cartesian and bispherical coordinates
\begin{align}
x &= \frac{a\sin\theta \cos\phi}{\cosh \eta-\cos\theta},\label{eq:x}\\
y &= \frac{a\sin\theta\sin\phi}{\cosh\eta-\cos\theta},\label{eq:y}\\
z &= \frac{a\sinh \eta}{\cosh \eta-\cos\theta},\label{eq:z}
\end{align}
where the radial bispherical coordinate is $\eta$,
the polar angular bispherical coordinate is
$\theta$
and the azimuthal angular bispherical coordinate is
$\phi = arctan(y/x)$. 
The half confocal length, $a$, determines the distance between
the bispherical coordinate poles, which is $2a$. When solving
problems between eccentric spheres, where one sphere is
inside a larger sphere, the confocal length is
\begin{equation}
a = \frac{\sqrt{R_i^4 + R_o^4 + \delta^4 - 2R_i^2R_o^2 - 2R_i^2\delta^2 - 2R_o^2\delta^2}}{2\delta},\nonumber
\end{equation}
where the eccentricity, $\delta$, is the distance between the inner
and outer spherical centers having, respectively, radii $R_i$ and
$R_o$. 

The ranges of $\eta,\theta$ and $\phi$ are $\eta\in[\eta_o,\eta_i]$, $\theta\in[0,\pi]$, $\phi\in[0,2\pi)$,
where
\begin{align}
\eta_o &= arcsinh(\frac{\sqrt{R_i^4 + R_o^4 + \delta^4 - 2R_i^2R_o^2 - 2R_i^2\delta^2 - 2R_o^2\delta^2}}{2\delta R_o}),\nonumber\\
\eta_i &= arcsinh(\frac{\sqrt{R_i^4 + R_o^4 + \delta^4 - 2R_i^2R_o^2 - 2R_i^2\delta^2 - 2R_o^2\delta^2}}{2\delta R_i}).\nonumber
\end{align}
In particular, for the symmetric spherical shell, as the eccentricity $\delta\to 0$, we have $\eta_o\to\infty$, $\eta_i\to\infty$ and $a\to\infty$. Subsequently, the bispherical polar coordinates reduce to the polar coordinates with $r =  \frac{a}{\cosh\eta-\cos\theta} = \frac{a}{\cosh\eta}$ and $\lim_{\eta\to\infty}\frac{\sinh\eta}{\cosh\eta} = 1$,
\begin{align}
x = \frac{a}{cosh\eta - cos\theta}\sin\theta\cos\phi = r\sin\theta\cos\phi,\nonumber\\
y = \frac{a}{cosh\eta - cos\theta}\sin\theta\sin\phi = r\sin\theta\sin\phi,\nonumber\\
(z-\frac{a\sinh\eta}{\cosh\eta}) = \frac{a}{\cosh\eta-\cos\theta}\frac{\sinh\eta}{\cosh\eta}\cos\theta = r\cos\theta,\nonumber
\end{align}
where $\frac{a\sinh\eta}{\cosh\eta}$ is the center of the spherical surface with fixed $\eta$, $r = R_i$ for $\eta = \eta_i$ and $r = R_o$ for $\eta = \eta_o$.

\section{Initial conditions and numerical methods in the Landau-de Gennes framework}\label{sec:b}
For the tetrahedron state on a symmetric shell in Fig. \ref{fig:TP}(a), we design an initial condition in terms of the director $\mathbf{n}$ as follows:
\begin{equation}
\mathbf{n} = \sin(\alpha)\mathbf{e}_{\theta} + \cos(\alpha)\mathbf{e}_{\phi} ,
\end{equation}
in the spherical coordinate with
\begin{equation}
\alpha = 
\begin{cases}
\pi + \phi &\text{if $\theta <\pi/4$},\\
-\pi/2 - \phi &\text{if $\theta > 3\pi/4$},\\
\pi/2 &\text{otherwise},
\end{cases}
\end{equation}
and use 
\begin{equation}
\mathbf{Q} = s_+(\mathbf{n}\otimes\mathbf{n} - \mathbf{I}/3),
\end{equation}
as an initial condition for the LdG numerical solver.
For the irregular tetrahedron state on an asymmetric shell in Fig. \ref{fig:TP}(c), with four defects concentrated near the south pole, we use the initial condition for the symmetric shell in terms of bispherical polar coordinate in Appendix \ref{sec:bispherical}, as the corresponding initial condition.

We numerically model the domain $\Omega$ using the bispherical coordinate system, $(\eta,\theta,\phi)$ in Appendix \ref{sec:bispherical}.
We expand the tensor function $\mathbf{Q}$ in terms of real spherical harmonics of $(\theta,\phi)$ and Legendre polynomials of $\zeta$ ($\zeta=2(\eta-\eta_o)/(\eta_i-\eta_o)-1$),
\begin{equation}\label{eq:expension}
q_i(\zeta,\theta,\phi) = \sum_{l = 0}^{L-1}\sum_{m = 1-M}^{M-1}\sum_{n=|m|}^{N-1} A_{lmn}^{(i)}Z_{lmn}(\zeta,\theta,\phi),
\end{equation}
where $N\geq M\geq L\geq 0$ specify the truncation limits of the expanded series, with
\begin{align}
&Z_{lmn}(\zeta,\theta,\phi) = P^l(\zeta)Y_{mn}(\theta,\phi),\\
&Y_{mn} = P_n^{|m|}(\cos\theta)X_m(\phi),\\
&X_m(\phi)=
\begin{cases}
\cos m\phi,  & \text{if $m\geq 0$}, \\
\sin|m|\phi, & \text{if $m<0$.}
\end{cases}
\end{align}
and 
$P_{n}^m(x)$ $(m\geq 0)$ are the normalized associated Legendre polynomials.
Using this series expansion, the LdG energy of $q_i$, $i = 1,\cdots,5$ is a function for the $5NML$ unknowns. Substituting \eqref{eq:expension} into the non-dimensionalized free energy \eqref{eq:energy-3d} and surface energy \eqref{eq:tangential},
we obtain a free energy as a function of these unknown
tensor order parameter elements, $A^{(i)}_{lmn}$. The
redefined free energy function is then minimized by using a standard optimization method, such as L-BFGS \cite{nocedal1999numerical} that treats the independent elements of tensor $A^{(i)}_{lmn}$ as variables.
The simulation results in Fig. \ref{fig:TP} are obtained by taking $(L, M, N) = (32, 64, 64)$.

\section{The boundary conditions of $D_{\phi,\theta}$ corresponding to a symmetric shell}\label{sec:bc}
The boundary conditions of $D_{\phi,\theta}$ corresponding to a symmetric shell is given by
\begin{align}
 &\alpha = \pi/2,\ \text{if}\ \phi = 0\ and\ \theta>\theta_1 + \epsilon,\label{eq:32}\\
 &\alpha = 0,\ \text{if}\ \phi = 0\ and\ \theta<\theta_1-\epsilon,\\
  &\alpha = \pi/2,\ \text{if}\ \phi = \pi/2\ \ and\ \theta<\theta_2 - \epsilon,\\
 &\alpha = 0,\ \text{if}\ \phi = \pi/2\ and\ \theta>\theta_2 + \epsilon,\\
 &\alpha = \phi,\ \text{if } \theta = 0,\\
 &\alpha = \pi/2-\phi\ \text{if } \theta = \pi,\\
 &\alpha = \pi/4-atan(\frac{\theta-\theta_1}{\phi})/2,\ \text{on $\partial D((0,\theta_1)$},\epsilon)\\
 &\alpha = \pi/4-atan(\frac{\theta-\theta_2}{\phi-\pi/2})/2,\ \text{on $\partial D((\pi/2,\theta_2),\epsilon)$}.\label{eq:39}
 \end{align}

 \section{The winding number of $\alpha = d\phi$ near the south pole $\theta = \pi$ is $d+1$.}\label{sec:d}
Substituting $\alpha = d\phi$ into \eqref{eq:alpha}, the corresponding director is
\begin{equation}\label{eq:n_d}
\mathbf{n} = \sin(\alpha)\mathbf{e}_{\theta}+\cos(\alpha)\mathbf{e}_{\phi} = \sin(d\phi)\mathbf{e}_{\theta}+\cos(d\phi)\mathbf{e}_{\phi}.
\end{equation}
Let us change back to Cartesian coordinates at $\theta = \pi$ with
\begin{align}
\mathbf{e}_{\theta} &= - \cos(\phi)\mathbf{e}_x - \sin(\phi)\mathbf{e}_y,\\
\mathbf{e}_{\phi} &= -\sin(\phi)\mathbf{e}_x + \cos(\phi)\mathbf{e}_y.
\end{align}
Substituting the above equations into \eqref{eq:n_d}, we have
\begin{align}
\mathbf{n} &= \sin(d\phi)(-\cos(\phi)\mathbf{e}_x-\sin(\phi)\mathbf{e}_y) \\
&+ \cos(d\phi)(-\sin(\phi)\mathbf{e}_x+\cos(\phi)\mathbf{e}_y) \\
&= - \sin((d+1)\phi)\mathbf{e}_x + \cos((d+1)\phi)\mathbf{e}_y,
\end{align}
the winding number of which is $d+1$ on $xy$-plane.

\bibliography{main}
\end{document}